\documentclass[letterpaper, 10 pt, journal, twoside]{IEEEtran}
\usepackage{amsmath,amsfonts}
\usepackage{algorithmic}
\usepackage{algorithm}
\usepackage{array}
\usepackage[caption=false,font=normalsize,labelfont=sf,textfont=sf]{subfig}
\usepackage{textcomp}
\usepackage{stfloats}
\usepackage{url}
\usepackage{verbatim}
\usepackage{graphicx}
\usepackage{cite}

\usepackage{amssymb}
\usepackage{booktabs}
\usepackage{hhline}
\usepackage{array}
\usepackage{booktabs}
\usepackage{amssymb}
\usepackage{pifont}
\usepackage{array}
\usepackage{tabularx}
\usepackage{graphicx}
\usepackage{booktabs}  % For \toprule, \midrule, \bottomrule
\usepackage{amssymb}   % For \checkmark and \times
\usepackage{makecell}  % For multi-line headers [1](@ref)
\usepackage{titlesec}
\usepackage[font=footnotesize,labelfont=bf]{caption}
\begin{document}

\title{ColonAdapter: Geometry Estimation Through Foundation Model Adaptation for Colonoscopy}

\author{Zhiyi Jiang$^{1}$, Yifu Wang$^{2}$, Xuelian Cheng$^{3}$$^{*}$, and Zongyuan Ge$^{4}$

% \thanks{Manuscript received: June 16, 2025; Revised: September 16, 2025; Accepted: October 3, 2025.}
% \thanks{This paper was recommended for publication by Editor Pascal Vasseur upon evaluation of the Associate Editor and Reviewers' comments.
% This work was supported by Jiangsu Department of Technology Natural Science Fund (Grants No: BK20250441) and Center of Excellence for Antimicrobial Therapeutics Discovery and Innovation (CEATDI)
% (Grants No: MSRI8002003).} %Use only for final RAL version
\thanks{$^{1}$Z. Jiang is with School of Computer Science and Engineering, Southeast University, China
        {\tt\footnotesize zyjiang97@outlook.com}}

\thanks{$^{2}$Y. Wang is Vetex Lab, Shanghai, China
        {\tt\footnotesize 1fwang927@gmail.com }}%

\thanks{$^{3}$X. Cheng is with Southeast University - Monash University Joint Graduate School, Suzhou 215123, China,  Monash Suzhou Research Institute, Monash University, Suzhou 215000, China, and also with Department of Data Science \& Artificial Intelligence (DSAI), Monash University, Clayton, VIC 3800, Australia
        {\tt\footnotesize  xuelian.cheng@monash.edu}}%

\thanks{$^{3} $Z. Ge is with Faculty of Information Technology,
Monash University, Clayton, VIC 3800, Australia
        {\tt\footnotesize  zongyuan.ge@monash.edu}}   

\thanks{Digital Object Identifier (DOI): see top of this page.}
}
% Paper headers
\markboth{IEEE Robotics and Automation Letters. Preprint Version. Accepted OCTOBER, 2025}
{JIANG \MakeLowercase{\textit{et al.}}: Geometry Estimation Through Foundation Model Adaptation for Colonoscopy}

\maketitle

\begin{abstract}
Estimating 3D geometry from monocular colonoscopy images is challenging due to non-Lambertian surfaces, moving light sources, and large textureless regions. While recent 3D geometric foundation models eliminate the need for multi-stage pipelines, their performance deteriorates in clinical scenes. These models are primarily trained on natural scene datasets and struggle with specularity and homogeneous textures typical in colonoscopy, leading to inaccurate geometry estimation. In this paper, we present ColonAdapter, a self-supervised fine-tuning framework that adapts geometric foundation models for colonoscopy geometry estimation. Our method leverages pretrained geometric priors while tailoring them to clinical data. To improve performance in low-texture regions and ensure scale consistency, we introduce a Detail Restoration Module (DRM) and a geometry consistency loss. Furthermore, a confidence-weighted photometric loss enhances training stability in clinical environments. Experiments on both synthetic and real datasets demonstrate that our approach achieves state-of-the-art performance in camera pose estimation, monocular depth prediction, and dense 3D point map reconstruction, without requiring ground-truth intrinsic parameters. https://github.com/JayJiang99/ColonAdapter
\end{abstract}


\section{Introduction}
\IEEEPARstart{C}{olorectal} cancer (CRC) is among the third most common type of cancer in the world, imposing a healthcare burden globally \cite{siegelColorectalCancer2023}. In the screening and treating of CRC, colonoscopy has been widely utilized as a gold-standard procedure \cite{rexQualityIndicators2015}. Despite its effectiveness, the colonoscopic procedure is subject to the experience of the clinician, as they have to screen a complex anatomical environment through monocular videos, which has a limited field of view and a lack of spatial information. To overcome these challenges, one promising approach is to estimate 3D organ geometry by dense reconstruction from monocular images \cite{zhangatemplate2021}. 
\begin{figure}[!t]
\centering
\includegraphics[width=3.5in]{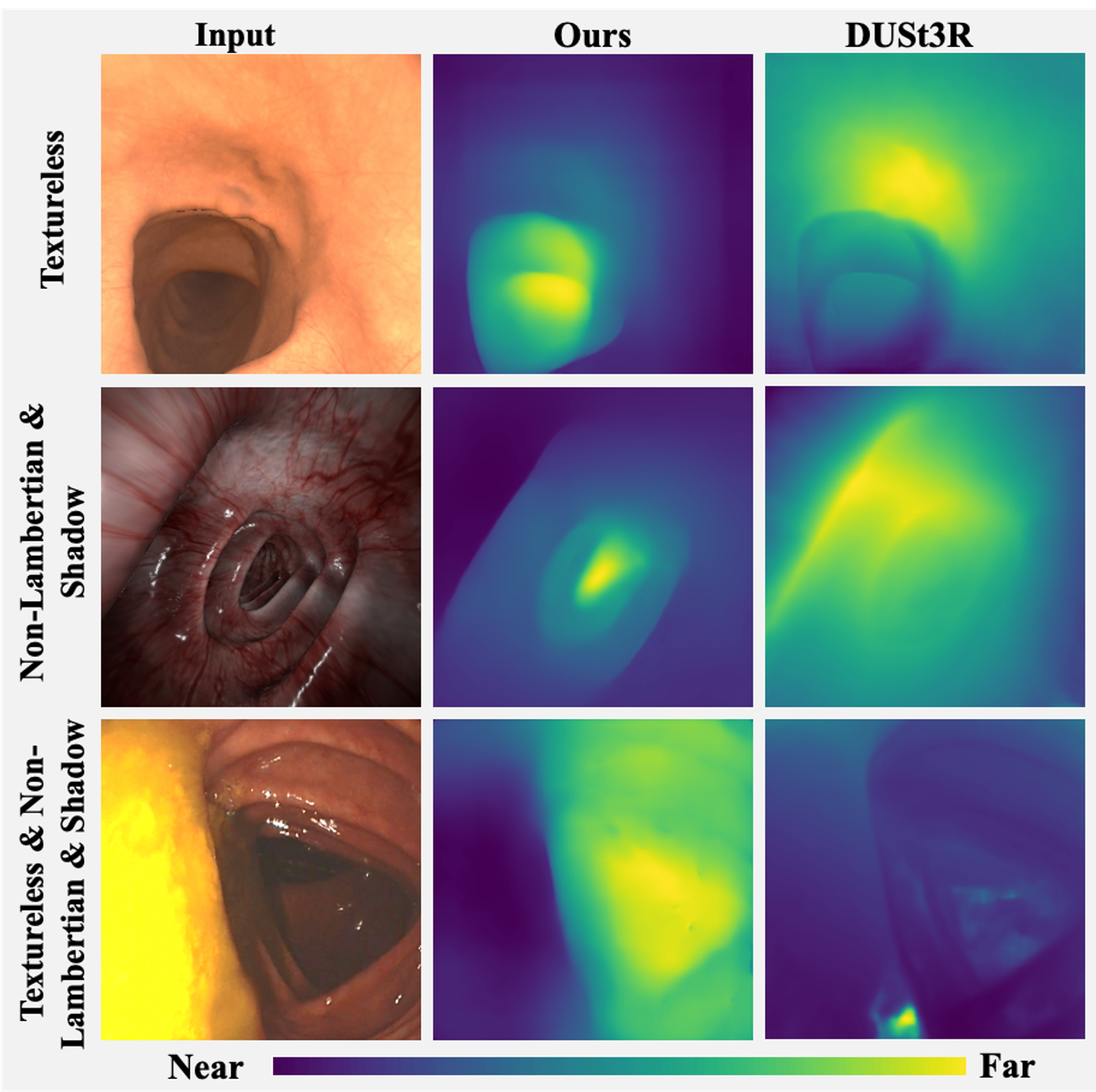}
\caption{Comparison of depth map estimations on colonoscopy images between a 3D geometric foundation model (DUSt3R \cite{wangDUSt3RGeometric3D2024}) and our proposed method. The top row shows how extensive textureless regions cause the model to misinterpret distant areas as close (and vice versa). The middle row highlights how shadows caused by moving light sources and complex anatomical structures lead DUSt3R to erroneously label them as distant. In contrast, the bottom row illustrates that our method accurately reconstructs scenes containing textureless surfaces, dynamic shadows, and non-Lambertian regions, where DUSt3R fails.}
\label{fig_incorrect-geometry}
\vspace{-7mm}
\end{figure}
Traditional feature-based reconstruction methods struggle in clinical scenarios because of the sparse and unevenly distributed key points in endoscopic images \cite{shaoSelfSupervisedMonocularDepth2021a}, resulting in poor reconstructions. To address these limitations, deep learning approaches that bypass handcrafted feature correspondences have gained traction for mapping and reconstruction in clinical settings. Some methods achieve dense reconstructions by incorporating learning-based depth and pose estimation within SLAM \cite{maRNNSLAMReconstructing3D2021a} or SfM \cite{recasensEndoDepthandMotionReconstructionTracking2021} frameworks. However, they typically assume known ground-truth intrinsic parameters, which are often unavailable or vary across devices. Additionally, their reliance on multi-stage reconstruction pipelines makes them prone to noise and error propagation \cite{wangDUSt3RGeometric3D2024}.

Recent advancements in 3D geometric foundation models \cite{wangDUSt3RGeometric3D2024} \cite{wangVGGTVisual2025} offer a promising solution by directly generating dense 3D point maps with 3D geometric information, eliminating the need for sub-modules and reliance on ground-truth intrinsic parameters. However, these models often struggle in challenging scenarios such as low-overlap or textureless regions, leading to degraded performance \cite{luLoRA3DLowRankSelfCalibration2024} \cite{ liMono3RExploitingMonocular2025}. LoRA3D \cite{luLoRA3DLowRankSelfCalibration2024} proposed a self-calibration technique to address these challenges, but it relies on the assumption that the overall scene geometry is relatively accurate. In natural scenes, this assumption is generally valid, as challenging regions typically occupy only a relatively small portion of the image, resulting in reliable scene reconstructions with only minor local errors. In clinical domains like colonoscopy, however, the situation is different: large textureless areas, complex anatomical structures, moving light sources, and non-Lambertian surfaces lead to inaccurate reconstruction of the entire scene, as shown in Fig.~\ref{fig_incorrect-geometry}. Consequently, existing methods are not suitable for addressing these substantial challenges for colonoscopy.

To adapt 3D geometric foundation models for colonoscopy, we propose a self-supervised fine-tuning strategy using only monocular videos, without camera information or ground-truth depth. Instead of relying on pixel-wise 3D point map losses, we demonstrate that photometric loss supervision suffices by leveraging the geometric priors of the foundation model. To enhance fine-detail recovery, we introduce a Detail
Restoration Module (DRM) that fuses fine details from an auxiliary convolutional encoder into the foundation model. This straightforward fusion strategy also ensures that the DRM can be seamlessly extended to other ViT-based geometric foundation models. 

Our main contributions are as follows: (1) Development of a self-supervised fine-tuning framework that adapts 3D geometric foundation models to colonoscopy scene through the supervision of photometric constraints. The confidence map is integrated into the photometric loss for training stability enhancement, and a geometric consistency loss is introduced to ensure coherent geometric predictions across frames. (2) Design of a Detail Restoration Module for ViT-based geometric foundation models, featuring an auxiliary CNN encoder and a feature fusion adapter for enhanced multi-level detail extraction and reconstruction accuracy in colonoscopy scenes. (3) Comprehensive evaluation on three colonoscopy datasets, demonstrating strong performance across multiple 3D vision tasks, including camera pose estimation, monocular depth estimation, and dense point map estimation.

\section{Related Work}

\subsection{Self-Supervised Learning}
Self-supervised learning has achieved notable success in a range of geometric perception tasks, including monocular depth prediction \cite{Godard_2019_ICCV}, multi-view monocular depth estimation \cite{watsonTemporalOpportunistSelfSupervised2021}, and structure-from-motion \cite{zhouUnsupervisedLearning2017}. These methods predominantly rely on photometric constraints, which perform well in natural scenes but are often violated in challenging environments, such as clinical scenarios. To address this, AF-SfMLearner \cite{shaoSelfSupervisedMonocularDepth2021a} introduced appearance flow to handle brightness inconsistencies, while MonoPCC \cite{wangMonoPCCPhotometricinvariant2025} proposed a photometric-invariant cycle consistency constraint. In parallel, geometric constraints have proven effective in addressing domain-specific challenges, as demonstrated by Wang et al. \cite{wangBridging2023}, who leverage inherent geometrical properties of satellite structures in their self-training approach for satellite pose estimation. Despite these advances, existing methods still rely on ground-truth camera intrinsics and predefined depth ranges during training. In contrast, our method processes consecutive frames as pair-view inputs and directly outputs scene geometry in the form of 3D point maps, from which both intrinsic parameters and camera poses can be derived.

\subsection{Vision Foundation Model Specialization}
Specializing vision foundation models through fine-tuning has become the standard approach for customizing pre-trained models to specific domains. For 3D geometric foundation models, several extensions have been proposed to enhance an existing architecture DUSt3R \cite{wangDUSt3RGeometric3D2024}. LoRA3D \cite{luLoRA3DLowRankSelfCalibration2024} applies low-rank adaptation (LoRA) \cite{hu2022lora} for efficient self-calibration.  Align3R \cite{luAlign3RAlignedMonocular2024} integrates monocular depth priors to enhance video depth estimation. However, these methods are not specifically designed for endoscopic or colonoscopic environments, which contain unique challenges such as large textureless regions, non-Lambertian surfaces, and moving light sources.

In endoscopy scenario, few works explored adapting depth anything model (DAM) with self-supervised framework. DARES \cite{zeinoddinDARESDepthAnything2024} applies LoRA to the DAM and joint training with a pose net. EndoDAC \cite{cuiEndoDACEfficientAdapting2024} introduced DV-LoRA and an extra intrinsic head for intrinsic estimation. With the intrinsic head, EndoDAC could be trained on any endoscopic videos. However, the separate models utilized for different sub-tasks making it vulnerable to noise and errors in each individual component. 

\subsection{Concurrent Work}
Endo3R \cite{Endo3RUnified} is a concurrent work that proposes a reconstruction framework for clinical scenarios and similarly incorporates optical flow into its loss formulation. A key distinction, however, lies in the supervision strategy: Endo3R relies on datasets with ground-truth depth or pseudo-labels generated by external video depth anything models, while our method is fully self-supervised and does not require ground-truth annotations or auxiliary depth models. Furthermore, our approach introduces an additional module to enhance the extraction of fine details, an aspect not addressed by Endo3R.

\section{Method}
\begin{figure*}[!t]
\centering
% \vspace{-7mm}
\includegraphics[width=7.0in]{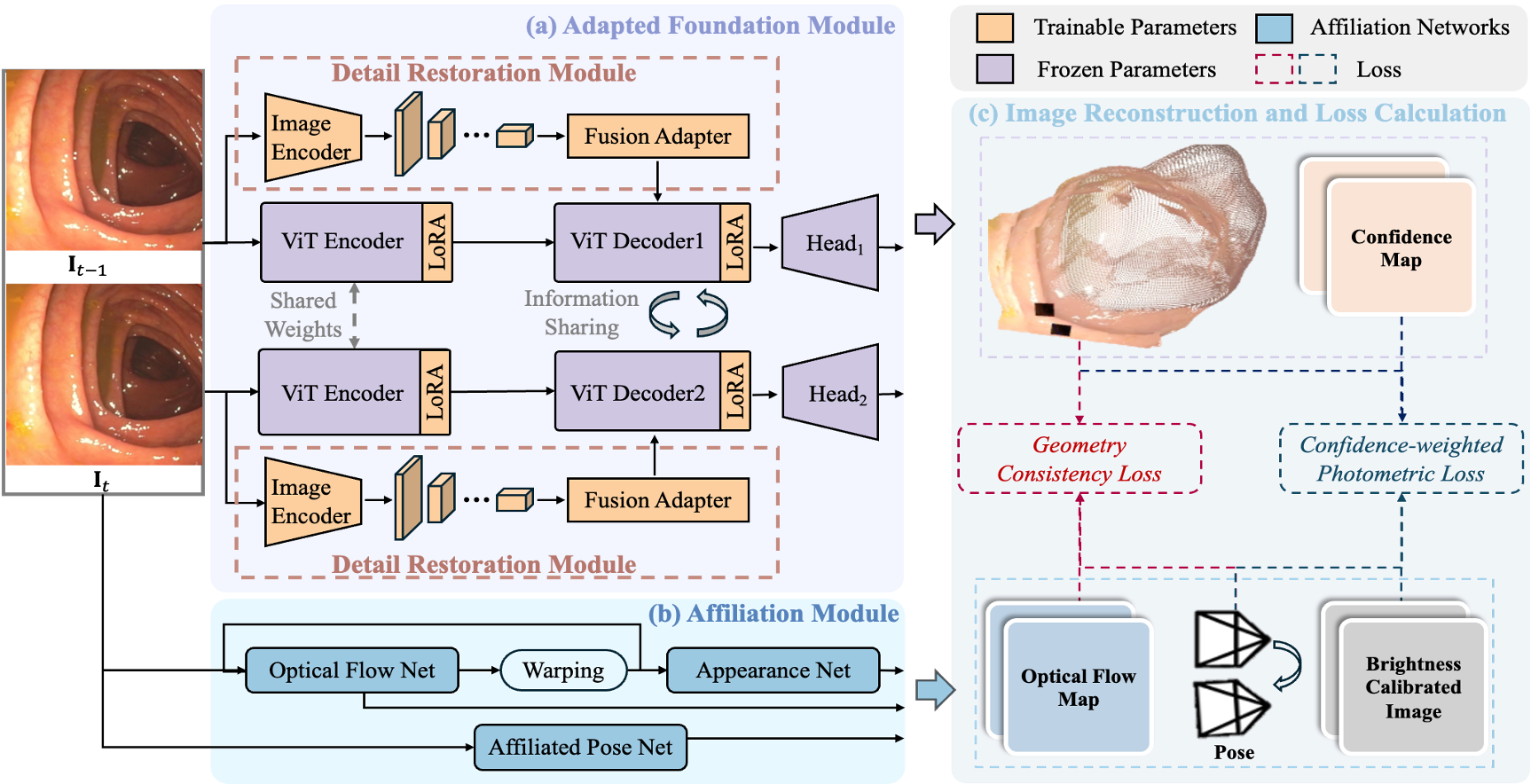}
\caption{The training pipeline of our proposed framework, consisting of an adapted foundation module (a), an affiliation module (b), and loss calculation (c). The adapted foundation module takes two input images and generates corresponding point maps along with their confidence maps. The affiliation module, used only during training, provides image brightness calibration, optical flow, and affiliated pose information. With the generated components from these two modules, we reconstruct image and calculate the losses.  During evaluation, we rely solely on the adapted foundation module to generate point maps, which are then used to derive camera information, including poses and intrinsic parameters.}
\label{fig_training_pipeline}
\vspace{-7mm}
\end{figure*}

We achieve geometry estimation using foundation model adaptation and further formulate such adaptation as 2D image reconstruction using a self-supervised framework. The overall training pipeline of the proposed framework is depicted in Fig.~\ref{fig_training_pipeline}, where DUSt3R is employed as the backbone foundation model and LoRA is used for fine-tuning. Section~A describes the details of overall framework. Section B introduces the Detail Restoration Module (DRM) for fine details enhancement. Section~C outlines the training objectives, including the confidence-weighted photometric loss and the geometry consistency loss.

\subsection{Overall Framework}
The framework main architecture, as illustrated in Fig.~\ref{fig_training_pipeline}, consists of adapted foundation module and affiliation module. The adapted foundation module generates the point map and calculates the intrinsic parameters using the point map. The affiliation module estimates the camera pose, calibrates brightness, and provides geometric consistency. The results of the two modules are subsequently input for image reconstruction and loss calculation.

\subsubsection*{(a) Adapted Foundation Module} 
% \subsubsection{Adapted Foundation Module}

The adapted foundation model primarily consists of a backbone foundation model with LoRA for fine-tuning and a Detail Restoration Module (DRM) for enhancing fine details, with further discussion of the DRM provided in Section B. The adapted foundation model takes two RGB images \( \mathbf{I}^t \), \(\mathbf{I}^{t-1}\) \(\in \mathbb{R}^{W \times H \times 3} \) as input and predicts corresponding point maps \( \mathbf{X}^{t; t} \), \( \mathbf{X}^{t-1; t} \) \(\in \mathbb{R}^{W \times H \times 3} \) , along with their associated confidence maps \( \mathbf{C}^{t; t} \), \( \mathbf{C}^{t-1; t} \) \(\in \mathbb{R}^{W \times H} \).  Here, \( \mathbf{X}^{t-1; t} \) represents the point map of frame \( t-1 \) expressed in the coordinate of frame \( t \). During inference, the model directly generates dense point maps that can be utilized to derive camera poses and intrinsic parameters. During training, we only utilize the point map \( \mathbf{X}^{t; t} \) and the derived intrinsic parameters to achieve the image reconstruction.

From the predicted point map, intrinsic parameters are estimated. This process assumes square pixels and centered principal points, thereby simplifying the task to estimating the focal length. Following the approach in \cite{wangDUSt3RGeometric3D2024}, the Weiszfeld algorithm is used to iteratively minimize the reprojection error:
\begin{align}
f^* = \arg\min_f \sum_{i} \left\| \mathbf{u}^t_i - f \cdot \mathbf{q}^t_i \right\|^2,
\label{eq:intrinsic_est}
\end{align}
where \( \mathbf{u}^t_i \) are the observed image points in image \( \mathbf{I}^t \) and \( \mathbf{q}_i \) is the corresponding normalized direction vector derived from point map \( \mathbf{X}^{t;t} \). This optimization typically converges within 10 iterations.

% \subsubsection{Affiliation Module}
\subsubsection*{(b) Affiliation Module} 

The affiliation module consists of an optical flow network, an appearance network, and an affiliated pose network, all of which are used exclusively during the training phase. Following the approach in \cite{shaoSelfSupervisedMonocularDepth2021a}, the optical flow and appearance networks are employed to address photometric inconsistencies caused by moving light sources and complex environments. Specifically, the optical flow network warps \( \mathbf{I}^{t-1} \) toward \( \mathbf{I}^t \). The resulting warped image is then input to the appearance network, together with \( \mathbf{I}^t \), to calibrate the brightness of \( \mathbf{I}^t \) and align it with \( \mathbf{I}^{t-1} \). This process calibrates the brightness in \( \mathbf{I}^t \) and generates  the calibrated image \( \tilde{\mathbf{I}}^t \).

 To provide pose information, we introduce an affiliated pose network to estimate the relative pose \( \mathbf{T}^{t \rightarrow t-1} \) \(\in SE(3) \) between the two frames. This choice is motivated by the fact that PnP algorithm used by \cite{wangDUSt3RGeometric3D2024} could fail and the algorithm accuracy also depends on the quality of the estimated confidence map. To improve generalizability, a pose network is utilized which relies solely on the image pair as input.
 
\subsubsection*{(c) Image Reconstruction and Loss Calculation} 

Given the estimated point map \( \mathbf{X}^{t;t} \) and relative pose \( \mathbf{T}^{t \rightarrow t-1} \), the target image \( \mathbf{I}^t \) is reconstructed by warping the source image \( \mathbf{I}^{t-1} \):
\begin{align}
\hat{\mathbf{I}}^t = \pi\bigl( K, \mathbf{T}^{t \rightarrow t-1}, \mathbf{X}^{t;t}, \mathbf{I}^{t-1} \bigr),
\label{eq:photometric_loss_orig}
\end{align}
where \( K \) \(\in \mathbb{R}^{3 \times 3} \) is the intrinsic matrix calculated from the focal length \( f \) \(\in \mathbb{R} \) and \( \pi(\cdot) \) denotes the re-projection operation. Here, \( \hat{\mathbf{I}}^t \) represents the reconstructed target image obtained from \( \mathbf{I}^{t-1} \). The reconstructed image is further utilized in the loss calculation.

The primary training objective is to minimize the difference between the reconstructed and target images, using photometric loss \cite{shaoSelfSupervisedMonocularDepth2021a} that combines the structural similarity index (SSIM) \cite{wang2004image} and pixel-wise differences. For each valid pixel \( i \), the photometric loss of frame \( t\) is defined as:
\begin{align}
l^{t}_{\text{photo}}(i) &= 
\alpha \cdot \frac{1 - \text{SSIM}(\tilde{\mathbf{I}}^t(i), \hat{\mathbf{I}}^t(i))}{2} \nonumber\\
&\quad + (1 - \alpha) \cdot \left\| \tilde{\mathbf{I}}^t(i) - \hat{\mathbf{I}}^t(i) \right\|_1,
\label{eq:photometric_loss}
\end{align}
where \( \tilde{\mathbf{I}}^t(i) \) represent the brightness-calibrated target image at valid pixel \( i \), which is determined using visibility masks \cite{shaoSelfSupervisedMonocularDepth2021a}. The balancing factor \( \alpha \) is set to 0.85. To facilitate the model training in colonoscopy, we further enhanced photometric loss with confidence and extra geometry consistency. The details for these two losses are expanded in Section~C. 

\subsection{Detail Restoration Module}

\begin{figure}[!t]
\centering

\includegraphics[width=3.5in]{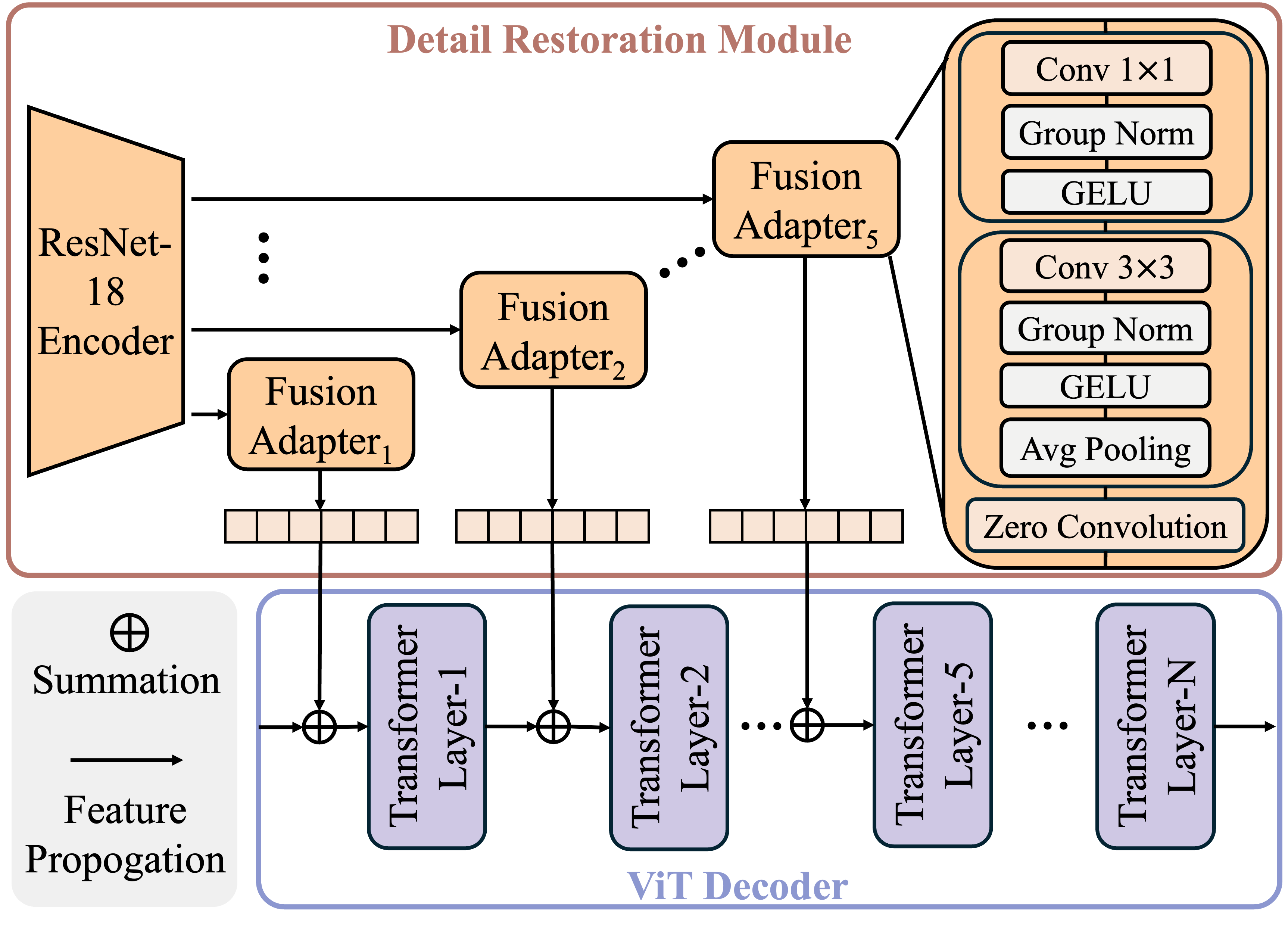}
\caption{Architecture of the proposed Detail Restoration Module. The ResNet-18 extracts multi-level features from the input images, which are then fused with ViT features through fusion adapters. The fused features are injected only into the first five layers of the ViT decoder to provide low-level information.}
\label{fig_drm_details}
\vspace{-7mm}
\end{figure}

As demonstrated in \cite{luAlign3RAlignedMonocular2024}, the DUSt3R generates only coarse 3D point maps. To enhance fine details, we incorporate fine-grained, high-frequency features extracted by a CNN encoder and fuse them with ViT features. This approach avoids relying on depth maps generated by monocular depth models \cite{luAlign3RAlignedMonocular2024}, which may suffer from scale ambiguity \cite{Wen_2025_CVPR}. For the CNN encoder, we adopt ResNet-18\cite{he2016resnet}, as it captures sufficient structural details in multi-level. Using heavier encoders yields only marginal improvements at the cost of significantly higher computational demands. The architecture of the Detail Restoration Module (DRM) is illustrated in Fig.~\ref{fig_drm_details}.

In the fusion of extracted CNN features and ViT features, we explored several strategies for feature fusion adapter, including: (1) a feature exchange block inspired by ViT-Adapter \cite{chenVisionTransformer2023} architectures to enable bidirectional information flow; (2) a Convolutional Block Attention Module (CBAM) \cite{yinCNNTransformerNetwork2023} , which introduces channel and spatial attention between two kinds of features; and (3) a zero convolution \cite{conf/iccv/ZhangRA23} with two blocks to match the channel and spatial dimensions of CNN and ViT features. Although the third method employs a simpler architecture with significantly fewer parameters, it achieves superior performance in our experiments, as shown in Table ~\ref{table:ablation_fusion}. Thus, it is selected as the final design for the feature fusion module. 

In the proposed DRM, a pair of input frames is processed by the ResNet-18 encoder (shared weights) to extract multi-level features. The multi-level features pass through five Fusion Adapters, each containing a channel projection block, a spatial projection block, and a zero convolution. The channel projection block utilizes \(1\times1\) convolution for channel alignment. The spatial projection block uses \(3\times3\) convolution and average pooling to refine features and match the spatial resolution of the ViT features. The aligned CNN features are subsequently fused with the ViT encoder outputs using the zero-convolution. Unlike \cite{luAlign3RAlignedMonocular2024}, which injects monocular priors into all transformer layers, we restrict our fusion to the first five layers of the ViT decoder. Following \cite{ranftlVisionTransformers2021}, deeper transformer layers capture high-level semantics, while shallow layers retain low-level details. Thus, we integrate low-level CNN features only into these shallow layers, enhancing detail reconstruction while preserving the high-level transformer representations.

\subsection{Training Objectives}
\subsubsection*{(a) Confidence-Weighted Photometric Loss} 

While this photometric loss supervises the estimated point maps, we observed that relying on it alone can slow down convergence. In some cases, it even degrades performance compared to the pre-trained foundation model when introducing the DRM, as shown in the fifth row of Table ~\ref{table:ablation_modules}. To address such issue, we introduce a confidence-weighted photometric loss that leverages per-pixel confidence maps predicted by the backbone model.

Confidence maps estimate the reliability of predictions at each pixel, allowing models to focus on reliable regions. This technique is used in 3D reconstruction frameworks like DUSt3R \cite{wangDUSt3RGeometric3D2024} and VGGT \cite{wangVGGTVisual2025}, enhancing robustness in occluded or low-texture areas. A direct approach is to adopt the confidence-aware loss from DUSt3R and replace the 3D regression loss with the photometric loss. However, using only the confidence-aware loss neglects low-confidence regions. To address this, we combine the photometric loss of frame \( t-1 \) with the confidence-aware loss of frame \( t \) to form the confidence-weighted photometric loss. This design ensures that low-confidence regions in frame \(t \) are supervised when warped to frame \(t-1 \) for photometric loss computation, without being discarded by the confidence map of frame \(t-1 \). In addition, such design complements the confidence-aware loss by providing stable gradient coverage, enhancing training robustness and optimization stability.

For the loss calculation, the model processes adjacent frames \(\mathbf{I}^t\) and \(\mathbf{I}^{t-1}\) bidirectionally, producing four point maps: \(\mathbf{X}^{t; t}\), \(\mathbf{X}^{t-1; t}\), \(\mathbf{X}^{t; t-1}\), and \(\mathbf{X}^{t-1; t-1}\). We use \(\mathbf{X}^{t; t}\) and its confidence map \(\mathbf{C}^{t; t}\) for the confidence-aware loss, and \(\mathbf{X}^{t-1; t-1}\) for the photometric loss:
\begin{align}
\mathcal{L}_{\text{conf-photo}} &= \lambda_{conf} \cdot \left( \frac{1}{N^{t}} \sum_{i=1}^{N^{t}} \mathbf{C}^t_i \cdot l^{t}_{\text{photo}}(i) - \beta \cdot \log(\mathbf{C}^t_i) \right) \nonumber\\
&\quad + \lambda_{photo} \cdot \left( \frac{1}{N^{t-1}} \sum_{i=1}^{N^{t-1}} l^{t-1}_{\text{photo}}(i) \right)
\label{eq:confidence_weighted_loss},
\end{align}
where \(( N^{t} , N^{t-1} )\) denote the numbers of valid pixels, and \( (\lambda_{conf} ,\lambda_{photo}) \) are weights for the confidence-aware and photometric losses. Incorporating the confidence map not only calibrates itself but also enhances downstream tasks that relies on confidence map such as PnP-based pose estimation and camera intrinsic estimation.

% \subsubsection{Geometry Consistency Loss}
\subsubsection*{(b) Geometry Consistency Loss}

The point maps from the adapted foundation module and the poses estimated by the affiliated pose network are inherently scale-invariant, which may lead to scale inconsistency. Furthermore, confidence-weighted photometric loss does not directly supervise \(\mathbf{X}^{t-1; t}\) and \(\mathbf{X}^{t; t-1}\).  To address both the scale inconsistencies and the lack of direct supervision for these point maps, we introduce a geometry consistency loss. This loss has two main components: one enforces scale consistency and geometric coherence across frames, while the other aligns the predicted camera poses scale from the affiliation module with the point maps scale.  This loss leverages the four point maps generated in the \(\mathcal{L}_{\text{conf-photo}}\) calculation.

For the first component, we align the scales of the point maps using the optical flow network within the affiliation module, which predicts the optical flows \(\mathbf{F}^{t \leftarrow t-1}\), \(\mathbf{F}^{t-1 \leftarrow t}\) \(\in \mathbb{R}^{W \times H \times 2} \) and the occlusion masks \(\mathbf{M}^{t \leftarrow t-1} \), \( \mathbf{M}^{t-1 \leftarrow t}\) \(\in \mathbb{R}^{W \times H } \). The optical flows warp the point maps \(\mathbf{X}^{t; t}\) and \(\mathbf{X}^{t-1; t-1}\) to produce \(\hat{\mathbf{X}}^{t-1; t}\) and \(\hat{\mathbf{X}}^{t; t-1}\). With the warped point maps and occlusion masks, the corresponding alignment term is defined as:
\begin{align}
\mathcal{T}_{\text{flow}} &= 
\mathbf{M}^{t \leftarrow t-1} \left\| \mathbf{X}^{t-1; t} - \hat{\mathbf{X}}^{t-1; t} \right\|_1 \nonumber\\
&\quad + \mathbf{M}^{t-1 \leftarrow t} \left\| \mathbf{X}^{t; t-1} - \hat{\mathbf{X}}^{t; t-1} \right\|_1.
\label{eq:flow_term}
\end{align}

For the second component, to align the estimated poses scale from the affiliation module with the point maps scale, we use the predicted poses \(\mathbf{T}^{t \rightarrow t-1}\) and \(\mathbf{T}^{t-1 \rightarrow t}\) to transform \(\mathbf{X}^{t-1; t}\) into the coordinate frames of \(t-1\) and \(t\), resulting in \(\hat{\mathbf{X}}^{t-1; t-1}\) and \(\hat{\mathbf{X}}^{t; t}\). The corresponding transformation term is defined as:
\begin{equation}
\mathcal{T}_{\text{pose}} = 
\left\| \mathbf{X}^{t-1; t-1} - \hat{\mathbf{X}}^{t-1; t-1} \right\|_1 +
\left\| \mathbf{X}^{t; t} - \hat{\mathbf{X}}^{t; t} \right\|_1.
\label{eq:pose_term}
\end{equation}

The final geometry consistency loss combines the two components, each weighted by separate factors \(\lambda_{\text{flow}}\) and \(\lambda_{\text{pose}}\):
\begin{equation}
\mathcal{L}_{\text{geo}} = 
\lambda_{\text{flow}} \cdot \mathcal{T}_{\text{flow}} + 
\lambda_{\text{pose}} \cdot \mathcal{T}_{\text{pose}},
\label{eq:geometry_loss}
\end{equation}
where \(\lambda_{\text{flow}}\) and \(\lambda_{\text{pose}}\) are the weights of the two components, respectively. 

\section{EXPERIMENTS, RESULTS, AND DISCUSSION}
\subsection{Experimental Setup}

\textbf{Simcol3D}:
Provided by the MICCAI 2022 EndoVis Challenge, SimCol3D \cite{RAU2024103195} is a synthetic colonoscopy dataset containing ground truth depth and pose. Virtual light sources are attached to the endoscope to simulate realistic illumination. For our experiments, we use data SyntheticColon I from SimCol3D, selecting 10 sequences for training and 3 for testing.

\textbf{Simulated Colonscopy Dataset (CSD)}:
The CSD  dataset is collected from a colonoscopy simulator\cite{zhangatemplate2021} offering a more complex environment with rich vascular textures. Three predefined paths are included; the first is used for testing, while the remaining two serve for training. This results in 8,246 training images and 2,034 testing images.

\textbf{EndoMapper}:
EndoMapper \cite{azagraEndomapperDatasetComplete2023} is a large-scale clinical dataset comprising 96 high-definition colonoscopy sequences, totaling over 24 hours of real-world video. Due to the absence of ground-truth depth, we extract representative frames from colonoscopy videos for qualitative evaluation only.

\textbf{C3VD}:
C3VD \cite{bobrow2023} is a phantom dataset acquired using an Olympus CF-HQ190L endoscope, containing 22 video sequences (10,015 images). For the quantitative evaluation of generalizability, We selected one representative sequence from each anatomical region for testing: Cecum (\texttt{cecum\_t2\_b}), Descending Colon (\texttt{desc\_t4\_a}), Sigmoid Colon (\texttt{sigmoid\_t3\_a}), and Transcending Colon (\texttt{trans\_t4\_a}).

\textbf{Implementation Details}:
All input images are resized to 224$\times$224 for training efficiency. Experiments are conducted using a single NVIDIA H100 GPU, with batch size 8. The learning rate is set to 1e-4. The CNN encoder (ResNet-18) is initialized with ImageNet-pretrained weights, while DUSt3R is initialized from its official pre-trained model.

\begin{table}[!b]
  \centering

  \caption{Quantitative camera pose estimation on various sequences of SimCol3D, evaluated using 10 random frames in accordance with \cite{zhangMonST3RSimple2024}. "G.I." indicates the use of ground-truth intrinsic parameters.}
  \label{table:pose-eval}
  \renewcommand{\arraystretch}{1.2}
  \begin{tabularx}{\linewidth}{l c >{\centering\arraybackslash}X >{\centering\arraybackslash}X >{\centering\arraybackslash}X}
    \toprule
    Method & {G.I.} & {ATE $\downarrow$} & {RPE Trans $\downarrow$} & {RPE Rot $\downarrow$} \\
    \midrule
    AF-SfMLearner & \checkmark & \underline{0.0067} & 0.1479 & 0.8518 \\
    Lite-Mono     & \checkmark & 0.0169 & 0.1438 & 0.5962 \\
    Monodepth2    & \checkmark & 0.0103 & 0.1459 & \underline{0.5679} \\
    DARES         & \checkmark & 0.0109 & 0.1473 & 0.9073 \\
    
    EndoDAC       & $\times$ & 0.0146 & \underline{0.1411} & {0.5787} \\
    \textbf{Ours} & $\times$ & \textbf{0.0062} & \textbf{0.0570} & \textbf{0.5573} \\
    \bottomrule
  \end{tabularx}
\end{table}

\begin{table}[!b]
  \centering
  \caption{Quantitative ego-motion comparison on the SimCol3D dataset. The ATE is averaged over all 5-frame snippets following\cite{shaoSelfSupervisedMonocularDepth2021a}.}
  \label{table:ego-motion}
  \renewcommand{\arraystretch}{1.2}
  \begin{tabularx}{\linewidth}{l c >{\centering\arraybackslash}X >{\centering\arraybackslash}X >{\centering\arraybackslash}X}
    \toprule
    {Method} & {G.I.} & {$\mathrm{ATE}_{s1}\downarrow$} & {$\mathrm{ATE}_{s2}\downarrow$} & {$\mathrm{ATE}_{s3}\downarrow$} \\
    \midrule
    AF-SfMLearner & \checkmark  & 0.2704 & 0.2710 & \underline{0.1591} \\
    Lite-Mono     & \checkmark & 0.3064 & 0.3114 & 0.1620 \\
    Monodepth2    & \checkmark & 0.2951  & 0.3143 & 0.1702 \\
    DARES         & \checkmark & 0.2678 & \textbf{0.2620} & 0.1594 \\
    
    EndoDAC       & $\times$ & 0.2711 & 0.2937 & 0.1596 \\
    \textbf{Ours} & $\times$ & \textbf{0.2654} & \underline{0.2660} & \textbf{0.1580} \\
    \bottomrule
  \end{tabularx}
\end{table}

\subsection{Camera Pose Estimation}

\begin{table*}[!t]
  \centering
  \caption{Monocular depth estimation on SimCol3D and CSD datasets. Best results are in \textbf{bold}, second-best are \underline{underlined}.}
  \label{table:mono-depth}
  \renewcommand{\arraystretch}{1.2}
  \setlength{\tabcolsep}{5pt}
  \begin{tabularx}{\textwidth}{l c *{5}{>{\centering\arraybackslash}X} *{5}{>{\centering\arraybackslash}X}}
    \toprule
    {Method} & {G.I.} &
    \multicolumn{5}{c}{{SimCol3D}} &
    \multicolumn{5}{c}{{CSD}} \\
    \cmidrule(lr){3-7} \cmidrule(lr){8-12}
    & & Abs Rel $\downarrow$ & Sq Rel $\downarrow$ & RMSE $\downarrow$ & RMSE log $\downarrow$ & $\delta \uparrow$ &
        Abs Rel $\downarrow$ & Sq Rel $\downarrow$ & RMSE $\downarrow$ & RMSE log $\downarrow$ & $\delta \uparrow$ \\
    \midrule
    AF-SfMLearner      & \checkmark & \underline{0.077} & 0.474 & 4.124 & \underline{0.111} & \underline{0.951} & \underline{0.071} & 0.208 & 2.565 & 0.110 & 0.950 \\
    Lite-Mono          & \checkmark & 0.092 & 0.708 & 3.976 & 0.123 & 0.937 & 0.080  &  \underline{0.207}   & \underline{2.428} & 0.117  & 0.943  \\
    
    Monodepth2         & \checkmark & 0.083 & 0.554 & \underline{3.861} & 0.112 & 0.950 & 0.100 & 0.668 & 3.208 & 0.129 & 0.915 \\
    DARES              & \checkmark & \underline{0.077} & \underline{0.439} & 4.458 & 0.121 & 0.944 & 0.116 & 0.616 & 3.339 & 0.157 & 0.881 \\
    ManyDepth         & \checkmark & 0.084 & 0.526 & 4.291 & 0.118 & 0.944 & 0.072 & \underline{0.207} & \textbf{2.347} & \underline{0.104} & \underline{0.959} \\
    EndoDAC            & $\times$     & 0.170 &	4.333 &	9.464 &	0.226	& 0.807     & 0.215   & 1.908   & 6.989      &  0.284     &  0.682   \\
    Dust3R      & $\times$   &  0.262 &	3.07	& 10.744 &	0.338	& 0.555	&	0.270 &	2.111 &	7.457	& 0.394 &	0.572       \\
    Ours  & $\times$    & \textbf{0.062} & \textbf{0.340} & \textbf{3.732} & \textbf{0.094} & \textbf{0.969} & \textbf{0.063} & \textbf{0.183} & 2.563 & \textbf{0.099} & \textbf{0.962} \\
    
    \bottomrule
  \end{tabularx}
\end{table*}

Camera pose estimation is evaluated on the SimCol3D dataset through two experiments. Following the protocol in \cite{zhangMonST3RSimple2024}, we randomly sample 10 frames from distinct segments of the colon and assess performance using Absolute Translation Error (ATE), Relative Translation Error (RPE-trans), and Relative Rotation Error (RPE-rot). All predicted poses are aligned to ground truth using Sim(3) Umeyama alignment.

We compare our method with other self-supervised joint depth-pose estimation method AF-SfMLeaner \cite{shaoSelfSupervisedMonocularDepth2021a}, Lite-Mono \cite{Zhang_2023_CVPR}, Monodepth2 \cite{Godard_2019_ICCV}, DARES \cite{zeinoddinDARESDepthAnything2024}, ManyDepth \cite{watsonTemporalOpportunistSelfSupervised2021}, and EndoDAC \cite{cuiEndoDACEfficientAdapting2024}. To ensure a fair comparison, all methods are trained on the same datasets as ours using the released code of authors. For EndoDAC, camera intrinsics learning is enabled during training. As shown in Table~\ref{table:pose-eval}, the proposed method outperforms the other methods across all metrics without known camera intrinsics in training. In particular, our method achieves significantly lower RPE values, benefiting from the simultaneous prediction of two point maps which enhances temporal consistency.

A second experiment is conducted using 3 full trajectories from the dataset in Table~\ref{table:ego-motion}. We follow the 5-frame evaluation protocol with ATE as the primary metric, consistent with \cite{shaoSelfSupervisedMonocularDepth2021a}. Three sequences are selected from different synthetic colons in SimCol3D. Results demonstrate that our approach achieves competitive performance compared to both supervised and self-supervised methods, highlighting its robustness across varied trajectories.

\begin{figure*}[!t]
\centering

\includegraphics[width=7.2in]{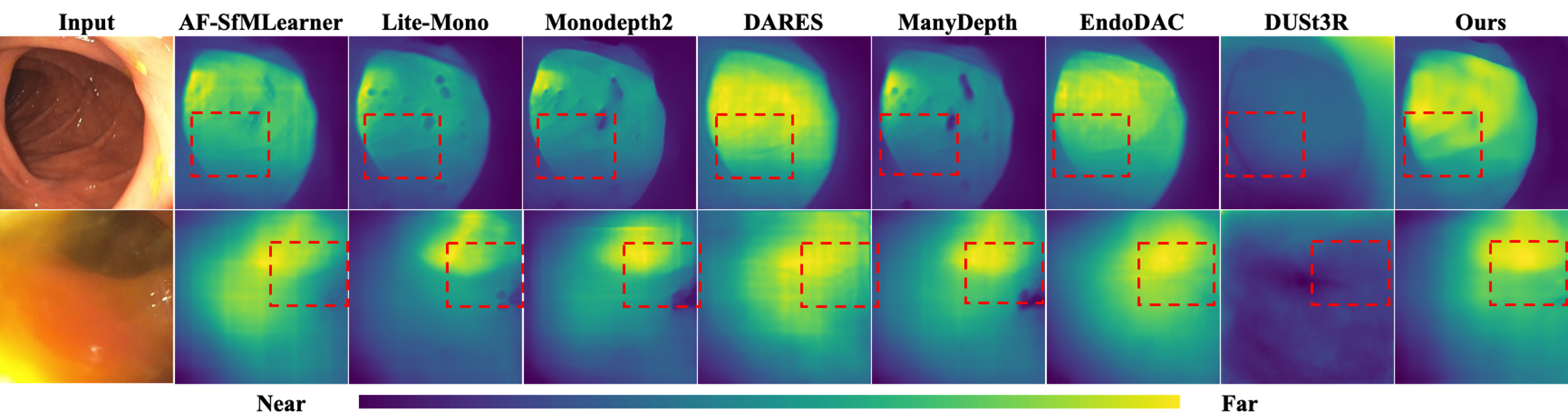}
\caption{Qualitative depth estimation results on the EndoMapper dataset. The top row highlights our method’s ability, enhanced by the integration of DRM, to capture fine structural details (highlighted with red box) that are missed by other approaches. The bottom row shows that even in the presence of unseen artifacts such as bubble, our method still predicts artifact-free geometry.}
\label{fig_realcolon}
\vspace{-7mm}
\end{figure*}

\subsection{Monocular Depth Estimation}

We provide quantitative evaluations on the SimCol3D and CSD datasets (Table~\ref{table:mono-depth}), comparing our method to single-view method \cite{shaoSelfSupervisedMonocularDepth2021a} \cite{Zhang_2023_CVPR} \cite{Godard_2019_ICCV} \cite{cuiEndoDACEfficientAdapting2024} \cite{zeinoddinDARESDepthAnything2024} and multi-view depth estimation method \cite{watsonTemporalOpportunistSelfSupervised2021}. Among them, EndoDAC \cite{cuiEndoDACEfficientAdapting2024} and DARES \cite{zeinoddinDARESDepthAnything2024} are depth foundation model adaptation approaches. Similar to camera pose estimation, the comparison methods are trained on the same datasets as ours using the released code of authors. We evaluated performance using five metrics: absolute relative error, square relative error, root-mean-square error, root-mean-square log error, and threshold accuracy. Notably, only EndoDAC does not require known camera intrinsics, while the other methods rely on ground-truth intrinsics and fixed depth ranges during training. Our method also avoids these dependencies, making the task more challenging, particularly on datasets like SimCol3D, which contain large textureless regions. Despite these challenges, our method achieves superior accuracy across most metrics.

For the CSD dataset, our RMSE is slightly higher than Lite-Mono and ManyDepth. This is because our model doesn't use predefined maximum depth ranges during training. Conventional monocular depth models benefit from these known limits, especially in tubular structures like the colon where inferring distant surfaces can be challenging. Since the CSD dataset mainly composed of tubular structure images with relatively large depth ranges, the performance of our model is impacted by the absence of depth range.

To evaluate generalizability, we conducted a quantitative analysis on the C3VD dataset and a qualitative analysis on the EndoMapper dataset using models trained on the SimCol3D dataset. As shown in Table~\ref{table:mono-depth-C3VD}, our method exhibits a slightly higher absolute relative error than DARES but outperforms DARES and other methods across the remaining metrics, indicating more stable and accurate predictions. Qualitative evaluation on the EndoMapper dataset shows improvements over DUSt3R (Fig.~\ref{fig_realcolon}), especially in challenging conditions like specularity and textureless regions. In the top row, methods that do not leverage foundation models (e.g., Monodepth2) exhibit noticeable artifacts. Although depth foundation model adaptation methods like DARES and EndoDAC reduce these artifacts, they still struggle to capture fine local structures (highlighted in red boxes). In the second row, the images include challenging conditions such as specularity, textureless region, and bubble. Our method accurately predicts the geometry without introducing artifacts, whereas other methods either introduce artifacts or fail to preserve fine details.

\begin{table}[!t]
  % \vspace{-5mm}
  \centering
  \caption{Monocular depth estimation on C3VD dataset. Best results are in \textbf{bold}, second-best are \underline{underlined}.}
  \label{table:mono-depth-C3VD}
  \renewcommand{\arraystretch}{1.2}
  \setlength{\tabcolsep}{5pt}
  \begin{tabularx}{\columnwidth}{l c *{5}{>{\centering\arraybackslash}X}}
    \toprule
    {Method} & {G.I.} &
    Abs Rel $\downarrow$ & Sq Rel $\downarrow$ & RMSE $\downarrow$ & RMSE log $\downarrow$ & $\delta \uparrow$ \\
    \midrule
    AF-SfMLearner      & \checkmark & 0.152 & 1.153 & 6.321 & \underline{0.190} & \underline{0.805} \\
    Lite-Mono          & \checkmark & 0.146 & 1.406 & 7.210 & 0.191 & 0.812 \\
    Monodepth2         & \checkmark & 0.143 & 1.248 & 7.241 & 0.188 & 0.820 \\
    DARES              & \checkmark & \textbf{0.134} & \underline{1.096} & 6.593 & 0.175 & 0.817 \\
    ManyDepth          & \checkmark & 0.190 & 1.960 & 8.641 & 0.260 & 0.716 \\
    EndoDAC            & $\times$   & 0.153 & 1.380 & 7.280 & 0.185 & 0.812 \\
    Dust3R             & $\times$   & 0.390 & 6.787 & 13.738 & 0.411 & 0.392 \\
    Ours               & $\times$   & \underline{0.139} & \textbf{0.956} & \textbf{5.592} & \textbf{0.175} & \textbf{0.832} \\
    \bottomrule
  \end{tabularx}

\end{table}

\subsection{Point Map Estimation}

We further evaluate our point map estimation results On SimCol3D. We add an extra completeness metric in addition to the three metrics: accuracy, square relative error, and RMSE log used in \cite{shaoSelfSupervisedMonocularDepth2021a}. Since the self-supervised approaches produce only depth maps, point clouds must be reconstructed by projecting depth predictions into 3D space using camera intrinsics. To ensure a fair comparison, we adopt the same scale normalization strategy used in depth evaluation to mitigate scale ambiguity.

For evaluation on the SimCol3D dataset, we select one method that requires ground-truth intrinsics (AF-SfMLearner), due to its strong performance in Table ~\ref{table:mono-depth}, and two methods that do not require ground-truth intrinsics (EndoDAC and DUSt3R). Quantitative results in Table~\ref{table:point-map} show that our method outperforms other three methods, demonstrating superior point map estimation accuracy even without access to ground-truth intrinsic parameters.

\begin{figure}[!t]
\centering

\includegraphics[width=3.35in]{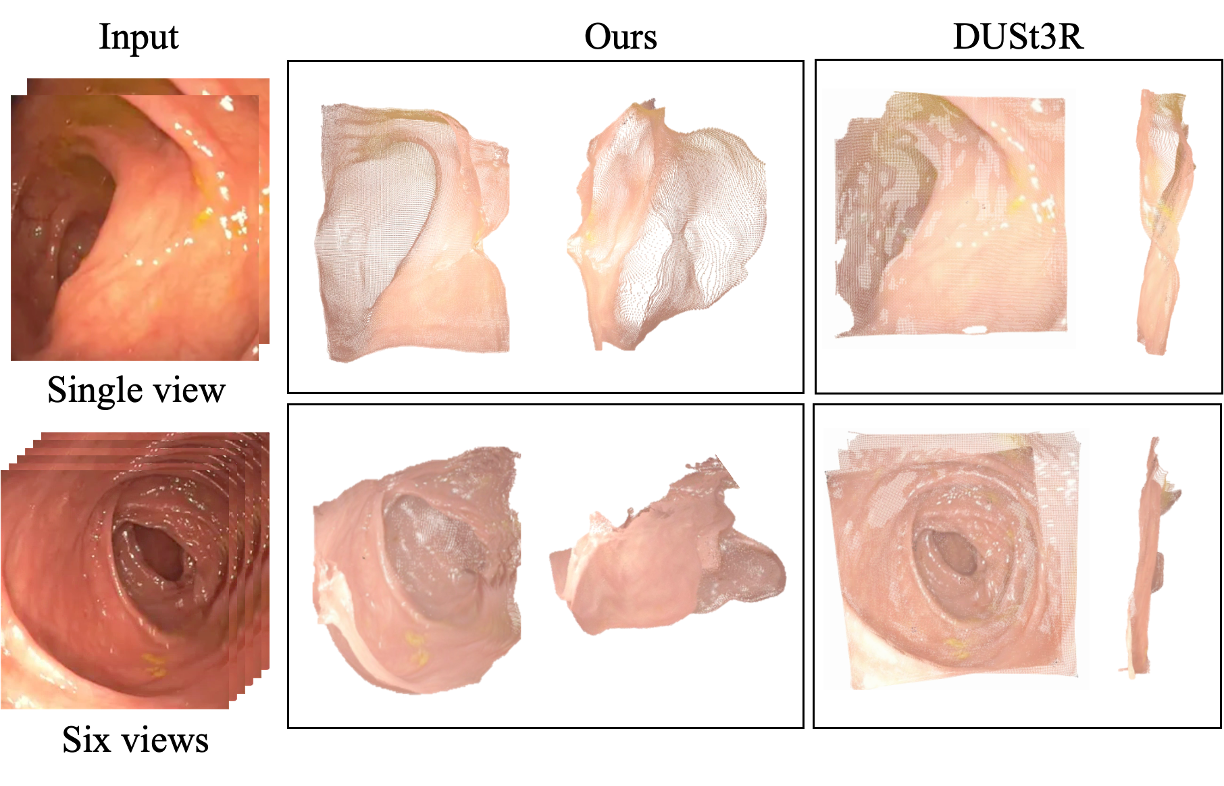}
\caption{Qualitative comparison of our predicted 3D point maps and the baseline DUSt3R on real colonoscopy images. In the top row, our method successfully recovers a 3D scene from two images containing textureless and non-Lambertian surfaces, while DUSt3R produces a distorted plane. In the bottom row, our method reconstructs the scene geometry, whereas DUSt3R predicts most of the region as a plane.}
\label{fig_pointmapestimation}
\vspace{-7mm}
\end{figure}

In addition to quantitative evaluation, we present two qualitative comparisons in Fig.~\ref{fig_pointmapestimation} to highlight the improvements of our model over DUSt3R on real colonoscopy images from the EndoMapper dataset. For reconstruction from more than two views, we follow DUSt3R to implement the global alignment. Our method produces high-quality reconstructions and generalizes well to challenging scenarios, including non-Lambertian surfaces and textureless regions.

\begin{table}[!b]
  \centering

  \caption{Point map estimation on SimCol3D}
  \label{table:point-map}
  \footnotesize  % Smaller font size
  \setlength{\tabcolsep}{3pt}  % Reduce column padding
  \renewcommand{\arraystretch}{1.1}
  \begin{tabular}{@{}l c c c c c@{}}  % Remove lateral padding
    \toprule
    {Method} & {G.I.} & {\makecell{Acc.\\$\downarrow$}} & {\makecell{Comp.\\$\downarrow$}} & {\makecell{SqRel\\$\downarrow$}} & {\makecell{RMSE\\log $\downarrow$}} \\
    \midrule
    AF-SfMLearner & \checkmark & \underline{1.779} & \underline{1.284} & \underline{0.032} & \underline{0.116} \\
    EndoDAC       & $\times$  & 19.610 & 8.566 & 2.463 & 0.368 \\
    DUSt3R        & $\times$  & 41.535 & 27.028 & 1.109 & 0.486 \\
    \textbf{Ours} & $\times$  & \textbf{1.742} & \textbf{1.182} & \textbf{0.028} & \textbf{0.111} \\
    \bottomrule
  \end{tabular}
\end{table}

\subsection{Ablation Study}
To assess the effectiveness of key components in our model design, we conduct ablation studies on the SimCol3D dataset, focusing on monocular depth estimation.

\begin{table}[!b]

  \centering

  \caption{Ablation study results on fusion strategy}
  \label{table:ablation_fusion}
  \renewcommand{\arraystretch}{1.2}
  \begin{tabularx}{\linewidth}{l c c c c c}
    \toprule
    {Method} & {Abs Rel $\downarrow$} & {Sq Rel $\downarrow$} & {RMSE $\downarrow$} & {RMSE log $\downarrow$} & {$\sigma \uparrow$} \\
    \midrule
    Fusion(1) & 0.075 & 0.528 & 4.666 & 0.114 & 0.944 \\
    Fusion(2) & 0.072 & 0.459 & 4.318 & 0.107 & 0.952 \\
    Fusion(3) & 0.062 & 0.340 & 3.732 & 0.094 & 0.969 \\
    \bottomrule
  \end{tabularx}
\end{table}

\begin{table}[!b]
  
  \centering
  \caption{Ablation study results}
  \label{table:ablation_modules}
  \renewcommand{\arraystretch}{1.2}
  \begin{tabularx}{\linewidth}{c c c c c c}
    \toprule
    \(\mathcal{L}_{\text{conf-photo}}\) & \(\mathcal{L}_{\text{geo}}\) & DRM & {Abs Rel $\downarrow$} & {RMSE $\downarrow$} & {$\sigma \uparrow$} \\
    \midrule
    \(\times\) & \(\times\) & \(\times\) & 0.204 & 8.384 & 0.699 \\
    \checkmark & \(\times\) & \(\times\) & 0.121 & 5.800 & 0.881 \\
    \checkmark & \checkmark & \(\times\) & 0.116 & 5.505 & 0.892 \\
    \checkmark & \(\times\) & \checkmark & 0.104 & 5.300 & 0.904 \\
    \(\times\) & \checkmark & \checkmark & 0.334 & 11.703 & 0.437 \\
    \checkmark & \checkmark & \checkmark & 0.062 & 3.732 & 0.969 \\
    \bottomrule
  \end{tabularx}
\end{table}

\subsubsection*{(a) Different Feature Fusion}

Different strategies for fusing CNN and transformer features yield varying results. The best performance is achieved using a lightweight sequential fusion strategy, as detailed in Table~\ref{table:ablation_fusion}. In contrast, more complex attention-based modules, such as feature exchange and CBAM, do not yield improvements. We hypothesize that while attention-based modules can refine representations in some scenarios, they may introduce noise or misaligned information that interferes with the pretrained transformer features.

% Furthermore, injecting CNN features into all transformer layers can worsen global structure recovery—likely due to the limited global context captured by shallow CNN features. Consequently, we restrict fusion to the initial transformer layers, ensuring a balance between leveraging local CNN features and preserving the global structural information captured by the transformer backbone.

\subsubsection*{(b) Impact of Loss Terms and Detail Restoration Module}

We evaluate the effects of the two proposed loss terms, confidence-weighted photometric loss and geometry consistency loss, alongside DRM. The first two rows in the table demonstrate that two losses enhance accuracy even without DRM. While the geometry consistency loss is not directly designed for accuracy improvement, it still contributes by ensuring both decoders (through cross-attention) are actively fine-tuned and complementary. 

After introducing DRM, these loss terms continue to improve performance. Notably, as shown in the fifth row, removing the confidence-weighted photometric loss leads to performance degradation, even dropping below the baseline foundation model. This underscores the stabilizing role of confidence weighting during training, as it guides the model to prioritize reliable regions.

\section{Conclusion}

In this paper, we introduce a self-supervised framework for fine-tuning 3D geometric foundation models in the colonoscopy domain. Following most foundation models, our method leverages point map representation to provide geometry information. The introduction of the Detail Restoration Module (DRM) enhances the extraction of fine details, which could also be applied to other ViT-based foundation models. Evaluations on both synthetic and real-world colonoscopy datasets demonstrate strong performance in pose estimation, monocular depth prediction, and dense point map reconstruction. However, utilizing DUSt3R as backbone introduces limitations, including high computational cost for long sequences, which restricts real-time scalability. In future work, we plan to integrate the adapted geometric foundation model into a SLAM framework to enable efficient and consistent 3D colon reconstruction.

% \section{References Section}
% You can use a bibliography generated by BibTeX as a .bbl file.
%  BibTeX documentation can be easily obtained at:
%  http://mirror.ctan.org/biblio/bibtex/contrib/doc/
%  The IEEEtran BibTeX style support page is:
%  http://www.michaelshell.org/tex/ieeetran/bibtex/
 
% \section{References}
\bibliography{citation}
\bibliographystyle{IEEEtran}

\end{document}